# Resonant properties of finite cracks and their acoustic emission spectra

V.V. Krylov

Department of Aeronautical and Automotive Engineering,
Loughborough University,
Loughborough, Leicestershire, LE11 3TU, UK

**Abstract**

In this paper, the acoustic emission accompanying the formation of brittle cracks of finite length is investigated theoretically using the approach based on the application of Huygens' principle for elastic solids. In the framework of this approach, the main input information required for calculations of acoustic emission spectra is the normal displacements of the crack edges as a function of frequency and wavenumber. Two simple approximate models defining this function are used in this paper for calculations of the acoustic emission spectra and directivity functions of a crack of finite length. The simplest model considers a crack that opens monotonously to its static value. The more refined model accounts for oscillations during crack opening and considers a crack of finite size as a resonator for symmetric modes of Rayleigh waves propagating along the crack edges and partly reflecting from the crack tips. Analytical solutions for generated acoustic emission spectra are obtained for both models and compared with each other. It is shown that resonant properties of a crack are responsible for the appearance of noticeable peaks in the frequency spectra of generated acoustic emission signals that can be used for evaluation of crack sizes. The obtained analytical results are illustrated by numerical calculations.

## 1. Introduction

Acoustic emission is a spontaneous radiation of elastic waves in solids and solid structures accompanying some irreversible processes, such as formation and development of brittle cracks, plastic deformation, dry friction, etc. [1-4]. In particular, the processes of formation and growth of cracks are causing radiation of elastic waves into the solid containing the crack and also into the surrounding space; the intensity of the associated sound radiation in the air is sometimes so great that it can be perceptible to the ear. The practical importance of this effect lies mainly in two aspects: it can be used for prediction of catastrophic failure of important engineering structures and also for research into solid state physics and fracture mechanics. The former aspect is especially important since radiated elastic waves provide information about developing cracks, which present the greatest danger.

In the present paper, the acoustic emission accompanying the formation of brittle cracks is investigated theoretically, using the approach earlier developed by the present author and his co-workers [5-9]. This approach is based on the application of Huygens' principle for elastic solids and on the use of suitable elastic Green's functions. In the framework of this approach, the main input information required for calculations of acoustic emission spectra is the normal displacements of the crack edges as a function of frequency and wavenumber. A precise description of this function for different situations of crack formation requires the solution of a complex problem of fracture mechanics, which is not always practical. For that reason, different approximations of this function can be used for practical calculations.

Two simple approximate models defining this function are used in this paper for calculations of the acoustic emission spectra and directivity functions of a crack of finite length opening under the impact of tensile stresses. The simplest model considers a crack that opens monotonously to its static value. The more refined model accounts for oscillations during crack opening and considers a crack of finite size as a resonator for symmetric modes of Rayleigh waves propagating along the crack edges and partly reflecting from the crack tips. Analytical solutions for generated acoustic emission spectra are obtained for both models and compared with each other. The obtained analytical results are illustrated by numerical calculations.

## 2. Outline of the theory

### 2.1 Statement of the problem

We assume that a crack is located in the interior or on the surface of an elastically stressed solid. For example, one can consider a stressed solid to be an infinite elastic half-space. The boundary conditions of zero normal stresses must be satisfied at the edges of the crack and on the free surface of the solid.

It is convenient to represent the whole problem of acoustic radiation by a crack developing in an elastic stressed medium as a superposition of the two problems, which can be done in linear approximation: 1) the static problem for the stressed solid without a crack (this problem is of no interest for the case under consideration); 2) the problem for an unstressed solid with a crack subjected to the nonzero normal stresses $n_i \sigma_{ij} = -n_i \sigma_{ij}^0$ applied to its edges in the absence of other sources of stresses. Here the quantities $\sigma_{ij}^0$ represent the stresses of the first problem calculated at the site of the crack (the stresses acting on the edges of the crack in the whole problem are equal to zero in this case, as expected). In what follows, we will be concerned only with the second problem, which is sufficient to describe the phenomenon of crack-induced acoustic emission.

### 2.2 Huygens' principle for elastic media

The theoretical approach considered in this paper is based on Huygens' principle for elastic solid media [10]. Applying it to the problem under consideration (for simplicity, we consider a two-dimensional case), we choose a closed contour $S$ running along the surface $z = 0$ around a cut indicating the possible path of crack propagation (this path is assumed to be known) and closed by a semicircle of infinite radius. Then the corresponding mathematical representation of Huygens' principle for time-harmonic fields, which is the analogue of the well-known Helmholtz integral theorem of classical acoustics, takes the form [5]:

$$u_m(\mathbf{r}) = \int_S [n_j \sigma_{ij} G_{im}(\mathbf{r},\mathbf{r}') - n_j c_{ijkl} u_i(\mathbf{r}') G_{im,k}(\mathbf{r},\mathbf{r}')] dS . \qquad (1)$$

Here the point of observation *r* lies inside the contour *S*, $u_m$ is the displacement vector, $n_i$ is the outward unit normal to the contour, $G_{im}(r, r')$ is the dynamic Green's tensor for an unbounded elastic medium, $c_{ijkl}$ are the elastic constants, and $G_{im,k} = \partial G_{im}/\partial x_k$. As usual, the summation over repeated indexes is assumed. The integral over the infinite semicircle vanishes, so that the integration in (1) is carried out only along the edges of the cut passing through the crack and along the boundary of the half-space [5]. Note that the first term in (1) vanishes in the integral along the free surface due to the stress-free boundary conditions on the surface $z = 0$. In addition, if to choose the Green's tensor that satisfies the stress-free boundary conditions on $z = 0$, then the second term in the integral over the free boundary vanishes as well, and the remaining integration should be carried out only over the crack area.

## 2.3 Plane crack in an infinite medium

In a number of situations, e.g., in the case of crack propagation in an unbounded medium, it is convenient to choose the Green's tensor in such a way that a certain type of specific boundary conditions is satisfied on the surface located between the edges of the crack. In what follows, we limit our consideration by a two-dimensional problem for a plane crack in an infinite elastic medium (Fig. 1), which is of fundamental importance.

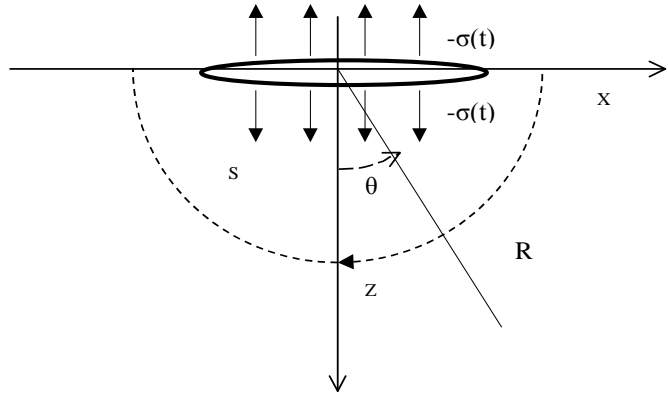

Fig. 1. Plane crack in an infinite elastic medium.

Let us introduce a Cartesian coordinate system with the *x* axis directed along the plane of the crack and with the *z* axis directed along the normal to it (see Fig. 1) and assume that an unbounded elastic medium is subjected to the action of normal tensile stresses $\sigma_{zz} = \sigma(t)$. According to the previous discussion, for the description of sound radiation, the action of these stresses is equivalent to the action of stresses $-\sigma(t)$ applied to the edges of the crack. Due to the symmetry of the problem about the *x* axis in the given case, it is sufficient to investigate the radiated field only in one half space, for example in the lower one. As a result, we have the mixed boundary conditions on the surface $z = 0$:

$$\sigma_{zz} = -\sigma(t) \text{ for } |x| < l; \quad \sigma_{zx} = 0 \text{ for } |x| \leq \infty \text{ and } u_z = 0 \text{ for } |x| > l. \quad (2)$$

To apply Huygens' principle to this problem we use the contour *S* as shown in Fig. 1. The general expression (1) then takes the form

$$u_m(x,z) = \int_{-\infty}^{\infty} [-\sigma_{zz}(x')G_{zm}(z, x-x') + c_{izkl}u_i(x')G_{lm,k}(z, x-x')]dx'. \qquad (3)$$

It is convenient to impose the following boundary conditions on Green's tensor $G_{lm}$:

$$\sigma_{zx} = 0, \quad u_z = 0, \quad |x| < \infty. \qquad (4)$$

It can be shown [5] that in this case the first term in equation (3) vanishes, and the second term contains only the normal component of crack edge displacement $u_z^0$. Expressing the convolution integral (3) in terms of the Fourier transforms, one can obtain

$$u_m(x,z,t) = \frac{1}{2\pi} \int_{-\infty}^{\infty}\int_{-\infty}^{\infty} [u_z^0(\omega,k)g_{zm}(z,\omega,k)]e^{ikx-i\omega t}d\omega dk, \qquad (5)$$

where $g_{zm}(z,\omega,k) = c_{zzkl}G_{lm,k}$ can be considered as a re-normalized Green's tensor satisfying the following boundary conditions:

$$\sigma_{zx} = 0 \quad \text{and} \quad u_z = \delta(x-x')\,\delta(t-t'). \qquad (6)$$

For isotropic solids, it is easier to find $g_{zm}(z,\omega,k)$ directly via the solution of the boundary problem defined by the homogeneous elastic wave equations subject to the boundary conditions (6). To obtain this solution, it is convenient to operate with Lame potentials $\varphi$ and $\psi$, rather than with displacements $u_m$. The displacement components $u_x$ and $u_z$ can be expressed in terms of the Lame potentials $\varphi$ and $\psi$ as follows:

$$u_x = \frac{\partial \varphi}{\partial x} - \frac{\partial \psi}{\partial z}, \qquad u_z = \frac{\partial \varphi}{\partial z} + \frac{\partial \psi}{\partial x}. \qquad (7)$$

Then, solving the well-known homogeneous wave equations for potentials $\varphi$ and $\psi$ subject to the boundary conditions (6), one can obtain, instead of the equation (5), the following expressions for the frequency spectra of the potentials $\varphi(\omega)$ and $\psi(\omega)$ [5, 8]:

$$\varphi(\omega) = \frac{i}{2\pi}\int_{-\infty}^{\infty} u_z^0(\omega,k)\frac{2k^2 - \omega^2/c_t^2}{(\omega^2/c_l^2 - k^2)^{1/2}\,\omega^2/c_t^2}e^{i(\omega^2/c_l^2 - k^2)^{1/2}z + ikx}dk, \qquad (8)$$

$$\psi(\omega) = \frac{i}{2\pi}\int_{-\infty}^{\infty} u_z^0(\omega,k)\frac{2k}{\omega^2/c_t^2}e^{i(\omega^2/c_t^2 - k^2)^{1/2}z + ikx}dk. \qquad (9)$$

Introducing a polar system of coordinates, $R$ and $\theta$, where angle $\theta$ is counted from the normal to the crack surface (see Fig. 1), one can express the radial and tangential components of displacements, $u_R$ and $u_\theta$ respectively, in the far field of radiation via the potentials $\varphi$ and $\psi$ defined by the equations (8) and (9):

$$u_R \cong \frac{\partial \varphi}{\partial R} = (i\omega/c_l)\varphi, \qquad (10)$$

$$u_\theta \cong \frac{\partial \psi}{\partial R} = -(i\omega/c_t)\psi. \tag{11}$$

Obviously, the displacements $u_R$ and $u_\theta$ in equations (10) and (11) are associated largely with longitudinal and shear elastic waves respectively. As it can be seen from equations (8) and (9), added by equations (10) and (11), the Lame potentials and displacements of radiated longitudinal and shear waves are fully determined by the spectrum $u_z^0(\omega,k)$ of the normal displacements of the crack edges, where the presence of the wave number $k$ in $u_z^0(\omega,k)$ accounts for current geometrical dimensions of the crack, in particular for its opening as well as for its growth from the nucleate stage to its macroscopic length.

The determination of the function $u_z^0(\omega,k)$ is a very complex and not completely solved problem of fracture mechanics. Therefore, simplified models of cracks opening and growth can be used, based on the available numerical or experimental data. One of the simplest models is the one where it is assumed that the crack initially grows instantaneously (at an infinite rate) as a slot to the length $2l$, and then its edges spread apart, eventually approaching their static position. This crack-opening process causes the acoustic emission from the crack, within the framework of the given model.

## 3. Calculations of acoustic emission spectra

In this section, we consider angular dependence (directivity functions) and frequency spectra of acoustic emission signals for two simple models of crack opening, which can be described by the appropriate functions $u_z^0$.

### 3.1 Monotonous opening of a crack of constant length

Let us consider the simplest model describing a monotonous opening of a crack with its tips' coordinates $l$ and $-l$ under the action of uniform tensile stresses $\sigma(t) = \sigma h(t)$, where $h(t)$ is Heaviside's step function. The following simple approximation for the function $u_z^0(x, t)$ can be used [11]:

$$u_z^0(x,t) = \begin{cases} u_z^0(t), & |x| \leq l \\ 0, & |x| > l \end{cases}, \tag{12}$$

where

$$u_z^0(t) = \begin{cases} st, & 0 \leq t \leq t_0 \\ s2l/c_l, & t > t_0 \end{cases}, \tag{13}$$

and $s = \sigma/\rho c_l$, $t_0 = 2l/c_l$. The rate $s$ of opening of the crack in equation (13) is determined as the result of dividing the applied tensile stress $\sigma$ by the wave impedance $\rho c_l$ of the medium in respect of the application of normal stresses. The crack opens up at this rate until it reaches the level $s2l/c_l$ corresponding to the well-known steady state (Westergaard's) solution. The real elliptical form of crack opening is disregarded in the above model.

The time form of crack opening $u_z^0(t)$ calculated according to equation (13) is shown in Fig. 2. The parameters used for calculations in Fig. 2 were as follows: the material – steel 1020, $c_l = 5893$ m/s, $c_t = 3240$ m/s, $\rho = 7820$ kg/m$^3$, $\sigma = 3 \cdot 10^8$ Pa, $2l = 0.02$ m.

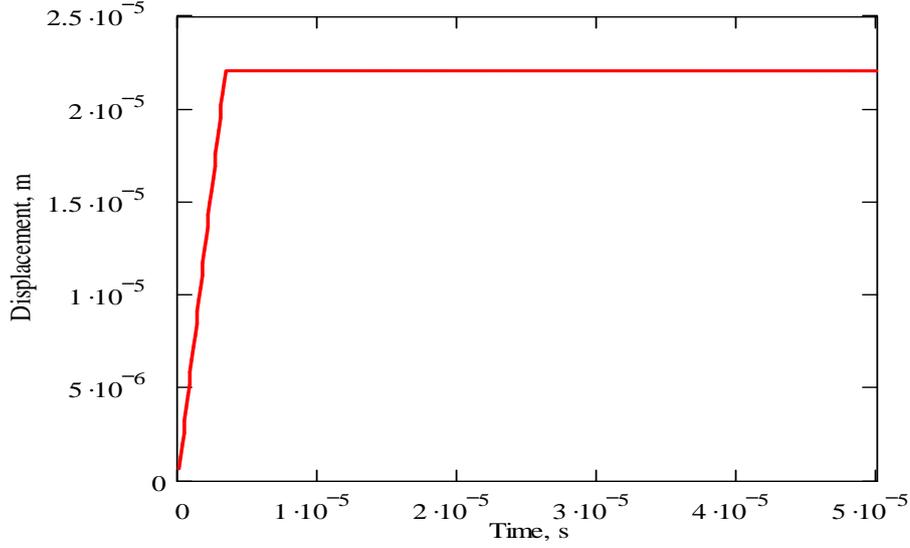

Fig. 2. The time form of monotonous crack opening $u_z^0(t)$ calculated according to equation (13).

Application of the double Fourier transform

$$u_z^0(\omega,k) = \frac{1}{2\pi} \int_0^\infty \int_{-l}^l u_z^0(t) e^{i\omega t - ikx} dt dx \qquad (14)$$

to the expressions (12) and (13) results in the following formula:

$$u_z^0(\omega,k) = u_z^{01}(\omega) u_z^{02}(k), \qquad (15)$$

where the function

$$u_z^{01}(\omega) = \frac{1}{2\pi} \int_0^\infty u_z^0(t) e^{i\omega t} dt = \frac{isl}{\pi \omega c_l} e^{i\frac{\omega}{c_l}l} \frac{\sin\frac{\omega}{c_l}l}{\frac{\omega}{c_l}l} \qquad (16)$$

describes the frequency spectrum associated with the time form $u_z^0(t)$ defined by equation (13), whereas

$$u_z^{02}(k) = \int_{-l}^l e^{-ikx} dx = 2l \frac{\sin kl}{kl}. \qquad (17)$$

accounts for the geometrical dimensions of the crack, as follows from equation (12). The absolute value of $u_z^{01}(\omega)$ as a function of frequency $f = \omega/2\pi$ is shown in Fig. 3. The parameters used for calculations in Fig. 3 were the same as those used in Fig. 2.

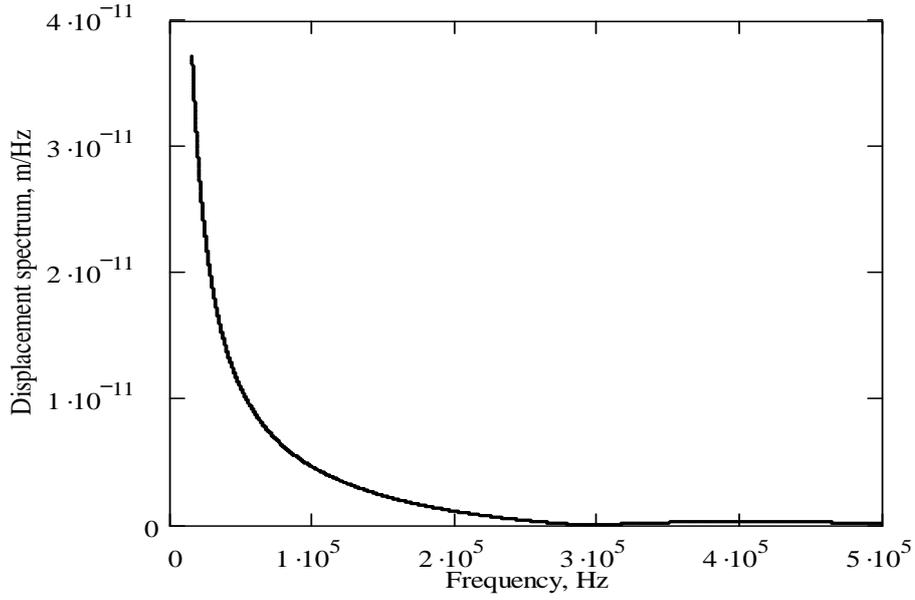

Fig. 3. Absolute value of $u_z^{01}(\omega)$ as a function of frequency $f = \omega/2\pi$ calculated according to equation (16).

Substituting formula (15) together with (16), (17) into the expressions (8) - (11) and applying the steepest descent method for the calculation of the integrals over $k$ in the far field of radiation, one can obtain the following expressions for the frequency spectra of the displacements $u_R$ and $u_\theta$:

$$u_R(R,\theta,\omega) = \frac{\omega}{2\pi c_l} u_z^{01}(\omega) \frac{2l \sin(\frac{\omega}{c_l} l \sin\theta)}{\frac{\omega}{c_l} l \sin\theta} \left(2\frac{c_t^2}{c_l^2}\sin^2\theta - 1\right)\left(-\frac{2\pi i}{(\omega/c_l)R}\right)^{1/2} e^{i\frac{\omega}{c_l}R}, \quad (18)$$

$$u_\theta(R,\theta,\omega) = -\frac{\omega}{2\pi c_t} u_z^{01}(\omega) \frac{2l \sin(\frac{\omega}{c_l} l \sin\theta)}{\frac{\omega}{c_l} l \sin\theta} \sin 2\theta \left(-\frac{2\pi i}{(\omega/c_t)R}\right)^{1/2} e^{i\frac{\omega}{c_t}R}. \quad (19)$$

Here $R$ and $\Theta$ are polar coordinates of the observation point: $x = R\sin\Theta$ and $z = R\cos\Theta$, and the function $u_z^{01}(\omega)$ is defined by formula (16). The analysis of the expressions (18) and (19) shows that the angular dependence (proportional to directivity function) of longitudinal waves (formula (18)) has a maximum in the normal direction to the crack (at $\Theta = 0$) for all spectral components, whereas shear waves (formula (19) are not generated in the normal direction at all, in agreement with the symmetry of the problem. The results of the numerical calculations of angular dependencies of longitudinal and shear waves for the crack under consideration, according to formulas (18) and (19), are shown in Fig. 4 for the values of $R$ and $f$ equal to 0.4 m and 70 kHz respectively. Other parameters are the same as in Figs. 2 and 3.

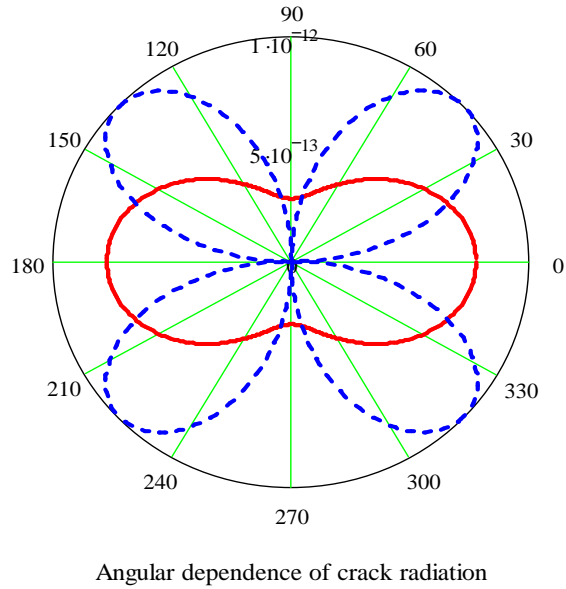

Angular dependence of crack radiation

Fig. 4. Angular dependencies of longitudinal waves (solid curve) and shear waves (dashed curve) radiated by an opening crack at frequency $f = 70$ kHz (the plane of the crack is in the vertical direction).

Angular dependencies of longitudinal and shear waves for the same crack calculated for the values of $R$ and $f$ equal to 0.4 m and 450 kHz respectively are shown in Fig. 5. Other parameters are the same as in Figs. 2-4.

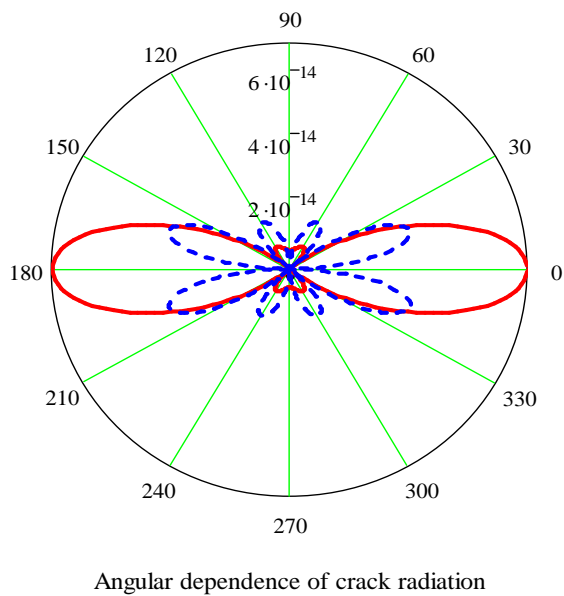

Angular dependence of crack radiation

Fig. 5. Angular dependencies of longitudinal waves (solid curve) and shear waves (dashed curve) radiated by an opening crack at frequency $f = 450$ kHz.

It can be seen from Fig. 5 that with the increase of frequency the role of radiated longitudinal waves becomes dominant.

The frequency spectra of radiated longitudinal and shear waves calculated using equations (18) and (19) are shown in Figs. 6 and 7 for the angle $\theta = \pi/12$ and for the values of crack length $2l = 0.02$ m and $2l = 0.04$ m respectively. The values of other relevant parameters are the same as those used for calculations in Figs. 4 and 5.

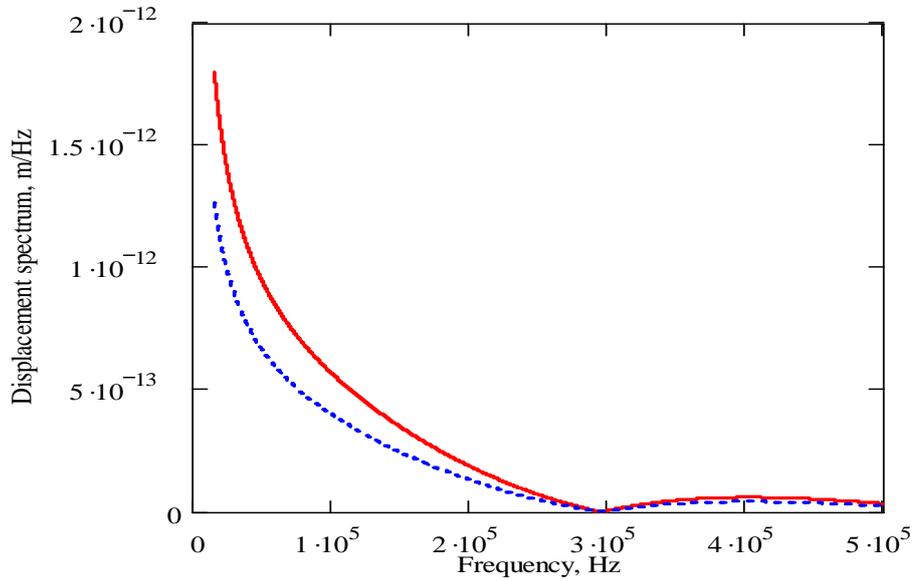

Fig. 6. Frequency spectra of longitudinal waves (solid curve) and shear waves (dashed curve) radiated by an opening crack of length $2l = 0.02$ m.

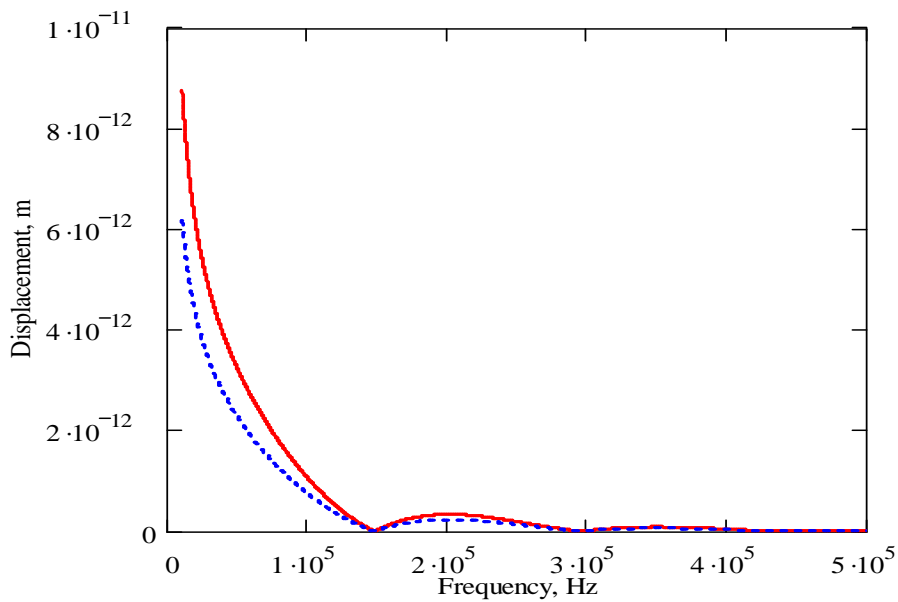

Fig. 7. Frequency spectra of longitudinal waves (solid curve) and shear waves (dashed curve) radiated by an opening crack of length $2l = 0.04$ m.

As it can be seen from Figs. 6 and 7, the frequency spectra of radiated longitudinal and shear waves are similar to each other. And there are the same zero values for both spectra defined by the equations  *sin(ωl/c_l) = 0* and  *sin[(ωl/c_l)sinθ] = 0*.

## 3.2 Crack opening with oscillations

A number of numerical calculations and experimental observations show that cracks of finite length are often opened to its static value not monotonously, as was considered in the previous section, but with a few decaying oscillations [12] responsible for the appearance of a noticeable maximum in the corresponding frequency spectrum of crack opening $u_z^{01}(\omega)$. The characteristic frequency $\omega_0$ of such oscillations depends on the length of the crack $2l$ and on one of the elastic wave velocities. It has been proposed earlier that physical mechanisms of such oscillations can be associated with elastic resonant phenomena taking place at the edges of a crack. In particular, one of the possible models of such resonant behaviour could be the one considering a crack as a resonator for Rayleigh surface waves propagating symmetrically along both crack's edges and reflecting from the crack's tips, loosing part of their energy to radiation of longitudinal and shear waves [8]. The resonant frequencies of such cracks has been predicted theoretically and compared with the experiments. The comparison of the predicted crack resonant frequencies with the results of experimental measurements of the acoustic emission spectra carried out for initiated cracks with different values of length $2l$ has shown their satisfactory agreement [8, 9]. In this section, we consider this issue in more detail, aiming not only at predicting resonant frequencies, but focusing our attention on the development of a semi-analytical model of crack opening and calculating the resulting acoustic emission spectra.

To analyse resonant properties of cracks of finite length it is important to know the reflection coefficients $R_0$ of symmetric modes of Rayleigh waves from the crack tips. On the basis of the numerical calculations [13] carried out for an elastic medium with Poisson's ratio $\nu = 0.25$, the complex value of the reflection coefficient from a crack tip $R_0$ can be estimated as $R_0 = 0.265 \exp(i\pi/2)$. It can be seen that the absolute value of the reflection coefficient, $|R_0| = 0.265$, is rather low. Substituting the phase of the above-mentioned reflection coefficient, $\eta = \pi/2$, into the equation for resonant frequencies of a crack of length $2l$ considered as a one-dimensional Fabry-Perot resonator,

$$\frac{\omega_0}{c_R} 2l + \eta = \pi n, \tag{20}$$

where $c_R$ is Rayleigh surface wave velocity in the material and $n$ is a mode number, one can easily express the values of resonant frequencies $\omega_0$. For example, for the lowest order mode ($n = 1$), it follows from equation (20) that

$$\omega_0 = \frac{\pi c_R}{4l}. \tag{21}$$

According to equation (21), for a finite crack of length $2l = 0.02$ m in 1020 steel, for which $c_R = 2997$ m/s, the value of the lowest resonant frequency (in Hz), $f_0 = \omega_0/2\pi$, is $3.747 \cdot 10^4$ Hz.

The quality factor $Q$ for a crack of length $2l$ considered as a one-dimensional Fabry-Perot resonator can be calculated as

$$Q = \frac{2\pi \cdot 2l}{\lambda_R(1-|R_0|^2)}, \qquad (22)$$

where $\lambda_R = 2\pi c_R/\omega_0$ is the Rayleigh wavelength at frequency $\omega_0$. Although the Poisson's ratio for 1020 steel, $\nu = 0.283$, is different from 0.25, for which the absolute value of the reflection coefficient $|R_0| = 0.265$ has been obtained, we will use this value of $|R_0|$ in equation (22) to estimate the quality factor $Q$ for the same crack of length $2l = 0.02$ m. The resulting value of $Q$ is 1.689, which means that a crack is a low quality resonator. The corresponding damping ratio $\zeta$ can be calculated as $\zeta = 1/2Q = 0.296$.

Based on the above, we can now modify the expression (13) for the time form function $u_z^0(t)$ to take into account crack oscillations in the process of its opening:

$$u_z^0(t) = \begin{cases} st, & 0 \le t \le t_0 \\ (s2l/c_l)[1 + p\sin(\omega_0(t-t_0))e^{-\zeta\omega_0(t-t_0)}], & t > t_0 \end{cases}, \qquad (23)$$

where $t_0 = 2l/c_l$, and $p$ is a non-dimensional semi-empirical coefficient describing the initial amplitude of crack oscillations about the established static value $s2l/c_l$. The time form $u_z^0(t)$ of the displacement of a finite crack of length $2l = 0.02$ m opening with oscillations calculated according to equation (23) for the values of the parameters $p = 0.6$, $f_0 = 3.747 \cdot 10^4$ Hz and $\zeta = 0.3$ is shown in Fig. 8. The values of other parameters are the same as in the previous figures.

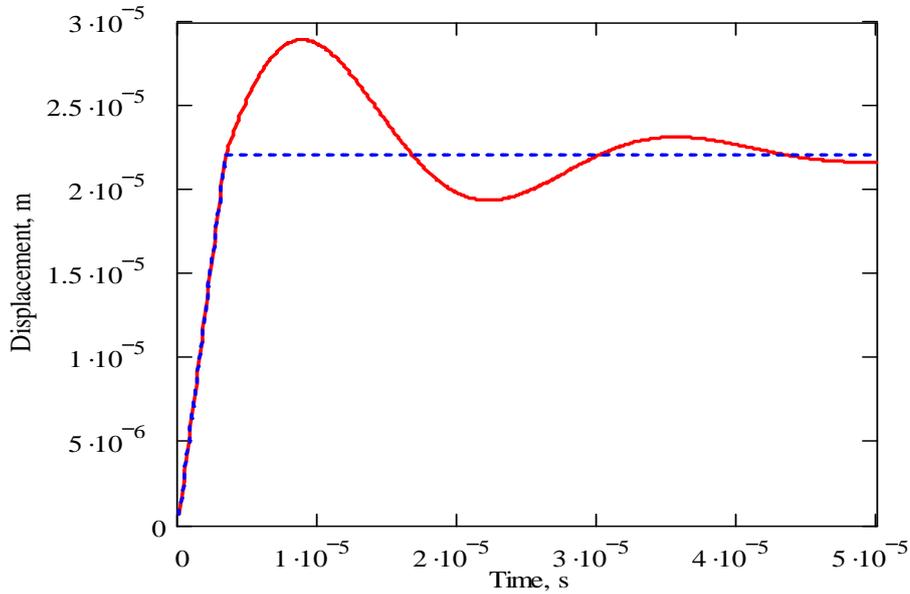

Fig. 8. The time form $u_z^0(t)$ of the displacement of a finite crack of length *2l* opening with oscillations (solid curve) about the established static value (dashed curve).

Taking the Fourier transform of the function $u_z^0(t)$ defined by equation (23), according to $u_z^{01}(\omega) = \frac{1}{2\pi}\int_0^\infty u_z^0(t)e^{i\omega t}dt$, results in the following expression for $u_z^{01}(\omega)$:

$$u_z^{01}(\omega) = \frac{isl}{\pi\omega c_l} e^{i\frac{\omega}{c_l}l} \frac{\sin\frac{\omega}{c_l}l}{\frac{\omega}{c_l}l} - \frac{slp}{\pi c_l} e^{i\frac{2\omega}{c_l}l} \frac{\omega_0}{\omega^2 - \omega_0^2 - \zeta^2\omega_0^2 + i2\zeta\omega_0\omega}. \qquad (24)$$

The behavior of $u_z^{01}(\omega)$ as a function of frequency $f = \omega/2\pi$ is shown in Fig. 9 in comparison with its behavior in the case of crack opening without oscillations. The parameters are the same as in Fig. 8.

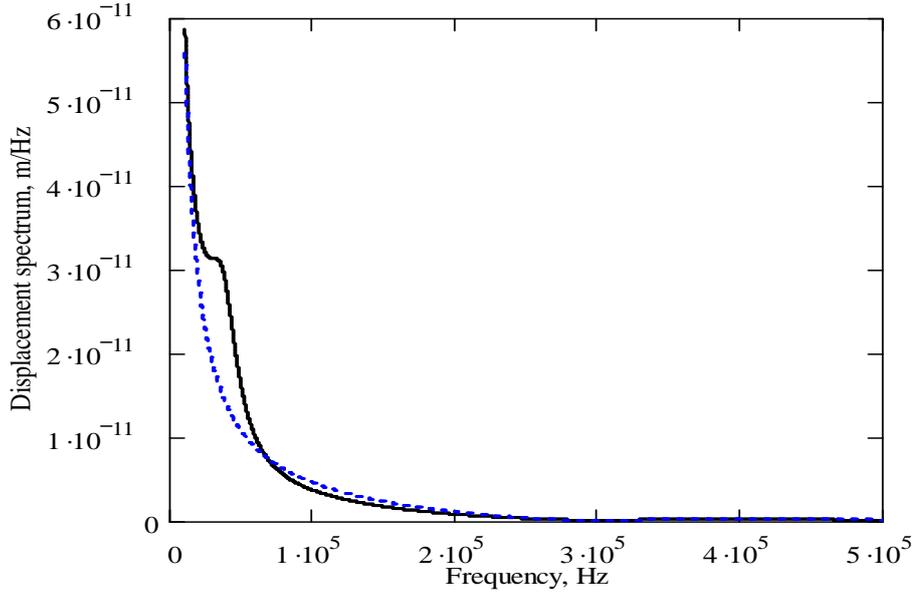

Fig. 9. The behavior of $u_z^{01}(\omega)$ as function of frequency in the cases of crack opening with oscillations (solid curve) and without oscillations (dashed curve).

To calculate the frequency spectra of the displacements $u_R$ and $u_\theta$ in the far field of radiation one can use the expressions (18) and (19) in which the function $u_z^{01}(\omega)$ is now defined by equation (24). The results of these calculations are shown in Fig. 10 for the crack length $2l = 0.02$ m and observation angle $\theta = \pi/12$. The values of the parameters associated with crack oscillations are $p = 0.6$, $f_0 = 3.747 \cdot 10^4$ Hz and $\zeta = 0.3$. Other parameters are the same as in the previous figures. It can be seen from Fig. 10 that in the case of crack opening with oscillations there are distinctive maxima at frequency $f_0$.

For comparison of the above results with the results obtained using a simpler model of a crack opening without oscillations ($p = 0$), Figure 11 shows the frequency spectra of the displacements $u_R$ only, for the cases of crack opening with oscillations and without oscillations. Again, the effect of oscillations is clearly seen in the appearance of the peak at frequency $f_0$ defined by equation (21). Note that this fact can be used for practical evaluation of sizes of developing cracks using the recorded acoustic emission spectra.

Obviously, in the framework of the applied model, crack oscillations do not influence angular dependencies (directivity functions) of radiated acoustic emission signals, which were considered in the previous section for a simpler model without oscillations. Therefore, angular dependencies are not discussed in the current section.

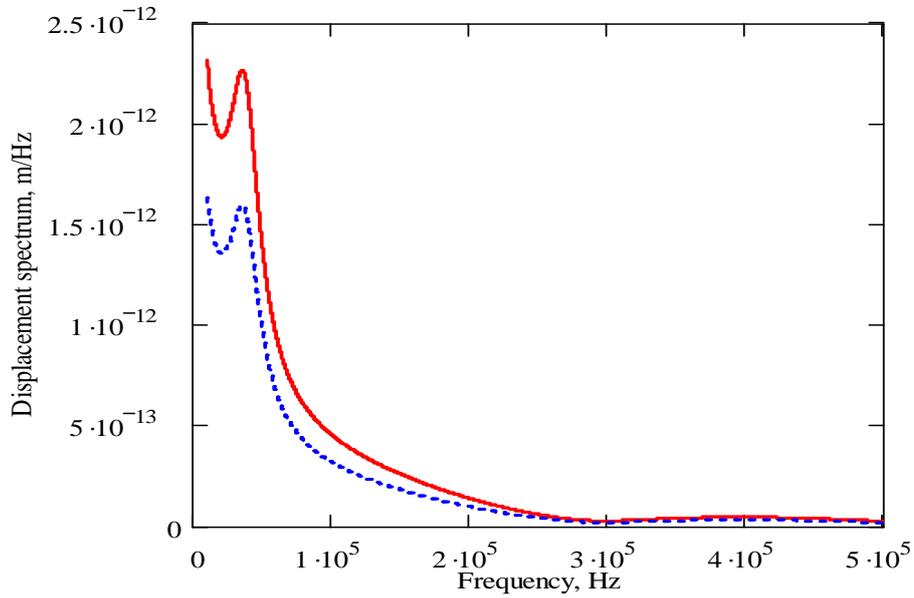

Fig. 10. Frequency spectra of longitudinal waves (solid curve) and shear waves (dashed curve) radiated by a crack of length $2l = 0.02$ m opening with oscillations.

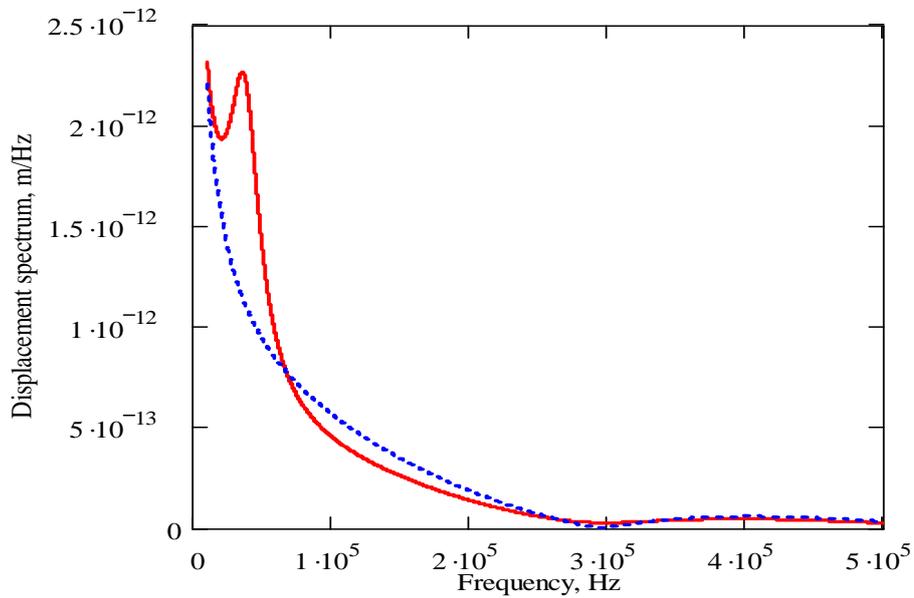

Fig. 11. Frequency spectra of longitudinal waves radiated by a crack of length $2l = 0.02$ m opening with oscillations (solid curve) and without oscillations (dashed curve).

## 3.3 Through cracks in thin plates

It should be noted that the two-dimensional problems of crack opening and the associated acoustic radiation considered in this paper are valid not only for unbounded media when the elastic fields are independent of the third coordinate $y$ (planar deformed state), but also in the case of through cracks in thin plates, i.e. if the plate thickness $h$ is much less that the wavelengths of longitudinal and shear waves (planar stressed state). In the case of thin plates subjected to tensile in-plane stresses, all formulas written in the previous sections remain valid, but the velocities $c_l$ and $c_t$ should be changed respectively to the velocity $c_L$ of the lowest order symmetric (quasi-longitudinal) Lamb mode of the plate (evaluated at $\omega \approx 0$), often called 'plate velocity', and to the velocity of the lowest order SH-mode $c_T$ (note that anti-symmetric Lamb modes, including the lowest order modes (flexural waves), are not generated by tensile stresses). The value of $c_L$ can be expressed in terms of the velocities in an unbounded medium as $c_L = 2c_t(1- c_t^2/c_l^2)^{1/2}$, whereas the velocity $c_T$ is simply equal to the velocity of shear waves, i. e. $c_T = c_t$.

The equivalent Poisson's ratio $v_1$ associated with the plate velocities $c_L$ and $c_T$ can be calculated according to the following well-known relation

$$v_1 = \frac{1 - 2(c_T^2 / c_L^2)}{2 - 2(c_T^2 / c_L^2)}. \tag{25}$$

This equivalent Poisson's ratio $v_1$ should be used for the calculation of the velocity $c_{R1}$ of quasi-Rayleigh waves propagating along edges of through cracks in plates, for example by means of the well-known approximate formula [14]:

$$c_{R1} = \frac{0.87 + 1.12 v_1}{1 + v_1} c_T. \tag{26}$$

Calculations carried out according to equations (25) and (26) for steel 1020 give the following values: $v_1 = 0.221$ and $c_{R1} = 2965$ m/s. This value of quasi-Rayleigh wave velocity should be used in equation (21), instead of $c_R$, for the calculation of crack resonant frequencies. For a through crack of length $2l = 0.02$ m, this gives the value of the lowest crack resonant frequency $f_{01} = 3.707 \cdot 10^4$ Hz. Using the above-mentioned plate-related velocities in equations (18) - (24) gives the frequency spectra of generated quasi-longitudinal waves and SH-waves in thin plates, $u_R(R,\theta,\omega)$ and $u_\theta(R,\theta,\omega)$ respectively. The results of the calculations for a through crack of length $2l = 0.02$ m in a thin plate made of steel 1020 are shown in Fig. 12 for angle $\theta = \pi/12$. Other parameters are the same as in the previous figures. As expected, the behaviour of $u_R$ and $u_\theta$ in this case is very similar to their behaviour in the case of planar deformed state (see Fig. 10).

In practical applications of acoustic emission in plates, measurements of displacements associated with radiated elastic waves are often taken in the normal direction to a plate surface (out of plane, rather than in plane). For time-harmonic waves, the normal displacements $u_n$ of a plate's surface are related to the in-plane displacements of quasi-longitudinal waves $u_R$ by the well-known equation

$$u_n = -\frac{vh}{2(1-v)} \frac{\partial u_R}{\partial R} = -\frac{i\omega}{c_L} \frac{vh}{2(1-v)} u_R, \tag{27}$$

where $h$ is thickness of the plate and $v$ is the Poisson's ratio of its material. It can be seen from equation (27) that $u_n(\omega)$ is proportional to $\omega u_R(\omega)$, which is why spectra of the normal displacements $u_n(\omega)$ are noticeably different from $u_R(\omega)$.

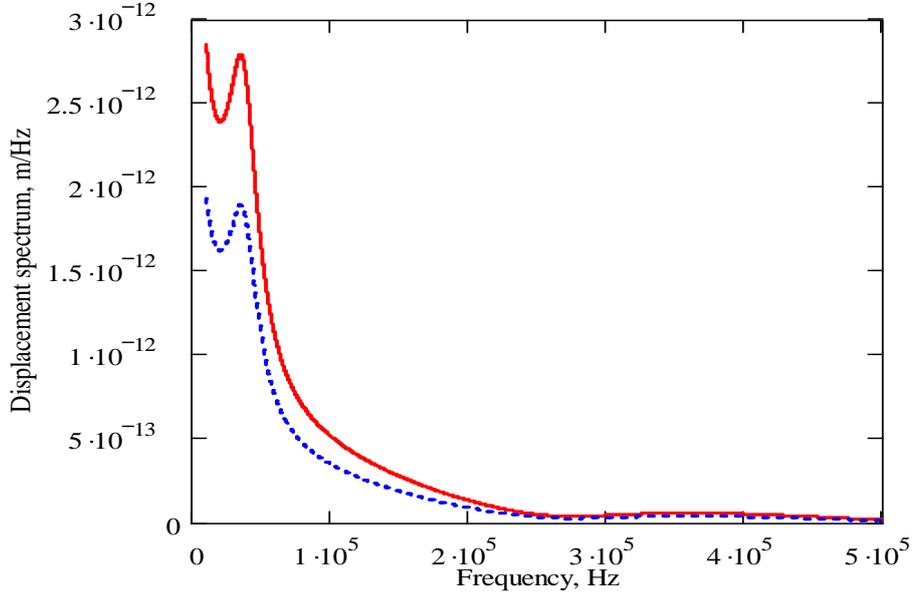

Fig. 12. Frequency spectra of quasi-longitudinal waves (solid curve) and SH-waves (dashed curve) radiated at angle $\theta = \pi/12$ in a thin plate by a crack of length $2l = 0.02$ m opening with oscillations.

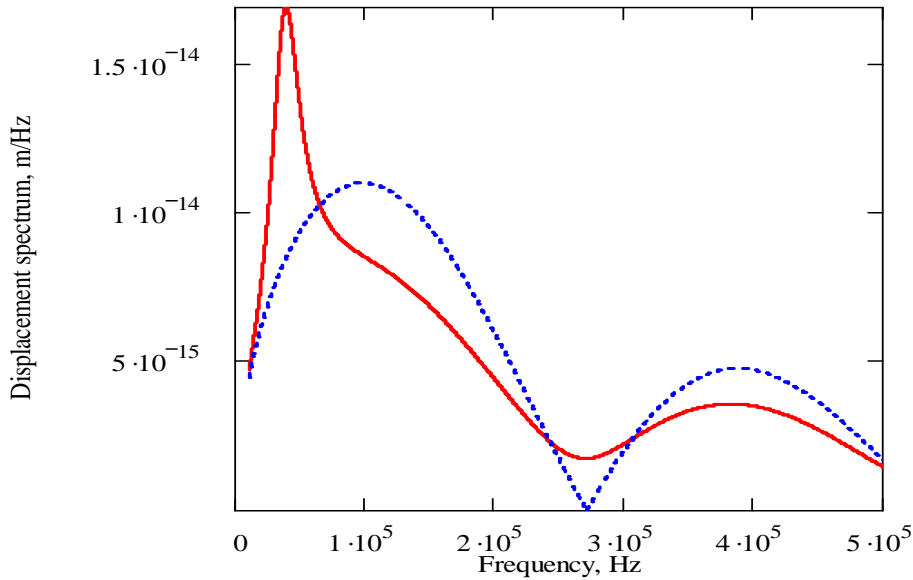

Fig. 13. Frequency spectrum $u_n(\omega)$ for a through crack of length $2l = 0.02$ m in a plate of thickness $h = 1$ mm opening with oscillations (solid curve) and without oscillations (dashed curve).

The results of the calculations of the frequency spectrum $u_n(\omega)$ for a through crack of length $2l = 0.02$ m opening with and without oscillations in a plate made of steel 1020 and having thickness $h = 1$ mm are shown in Fig. 13. The Poisson' ratio of the plate material used in equation (27) for this case is $v = 0.283$, and the plate velocity is $c_L = 5413$ m/s. Other parameters are the same as in Fig. 12. It can be seen from Fig. 13 that, as discussed, the spectrum $u_n(\omega)$ looks noticeably different from the associated spectrum $u_R(\omega)$ shown in Fig. 12. In particular, the maximum at the crack resonant frequency becomes more pronounced.

## 4. Conclusions

In this paper, the acoustic emission spectra associated with the formation of two-dimensional brittle cracks of finite length have been investigated theoretically using the approach based on the application of Huygens' principle for elastic solid media. As a result of the suitable choice of Green's tensor, the main input information required for calculations of the acoustic emission spectra from a crack developing in an unbounded elastic medium is the normal displacement of the crack edges as a function of frequency and wavenumber.

Two simple approximate models defining the normal displacement of the crack edges have been used in this paper for calculations of the acoustic emission spectra and directivity functions of cracks of finite length opening under the impact of tensile stresses. The simplest model considers a crack that opens monotonously to its static value. The more refined model accounts for oscillations during crack opening and considers a crack of finite size as a resonator for symmetric modes of Rayleigh waves propagating along the crack edges and partly reflecting from the crack tips. Analytical solutions for generated acoustic emission spectra have been obtained for both models and compared with each other. In particular, it has been shown that, in the case of crack opening with oscillations, the acoustic emission spectra of radiated longitudinal and shear waves have the distinctive maxima at frequencies defined by resonant frequencies of cracks considered as resonators for symmetric modes of Rayleigh waves. This fact can be used in practice for evaluations of sizes of developing cracks using their acoustic emission spectra. The obtained analytical results have been illustrated by numerical calculations.

The above-mentioned two simple models of crack opening have been also applied to the analysis of acoustic emission signals radiated by through cracks in thin plates. For in-plane displacements of radiated waves, the calculated spectra are very similar to those obtained for two-dimensional unbounded problems. For normal displacements of plate surfaces, the calculated spectra are noticeably different, and the maxima at crack resonant frequencies are more pronounced.

## References

[1] C.B. Scruby, An introduction to acoustic emission, Journal of Physics E: Scientific Instruments, 20, 946-953 (1987).
[2] M. Ohtsu and K. Ono, The generalized theory and source representation of acoustic emission. Journal of Acoustic Emission, 5, 124–133 (1986).
[3] K. Ohno and M. Ohtsu, Crack classification in concrete based on acoustic emission, Construction and Building Materials, 24, 2339–2346 (2010).

[4] M.G.R. Sause and S. Richler, Finite element modelling of cracks as acoustic emission sources, Journal of Nondestructive Evaluation, 34:4 (13 pp) (2015).

[5] V.V. Krylov, Radiation of sound by growing cracks. Soviet Physics - Acoustics, 29(6), 468-472 (1983).

[6] V.V. Krylov and E.P. Ponomarev, Acoustic emission accompanying the onset of surface microcracks. Soviet Physics - Acoustics, 31(2), 122-125 (1985).

[7] V.V. Krylov and E.P. Ponomarev, Model study of the acoustic emission of surface microcracks. Soviet Physics - Technical Physics, 30(8), 958-959 (1985).

[8] V.V. Krylov and E.P. Ponomarev, Acoustic emission spectra from the formation of through cracks in glasses. Soviet Physics - Acoustics, 32(5), 386-389 (1986).

[9] V.V. Krylov, P.S. Landa and V.A. Robsman, Model of the evolution of acoustic emission as the randomization of transient processes in coupled nonlinear oscillators. Acoustical Physics, 39(1), 55-61 (1993).

[10] Y.-H. Pao and V. Vararharajulu, Huygens' principle, radiation conditions, and integral formulas for the scattering of elastic waves. Journal of the Acoustical Society of America, 59(6), 1361-1371 (1976).

[11] L.A. Maslov, Model of a fracture as an emitter of elastic vibrations. Journal of Applied Mechanics and Technical Physics, 17(2), 274-279 (1976).

[12] V.Z. Parton and V.G. Boriskovskii, Dynamical Fracture Mechanics (in Russian), Mashinostroenie, Moscow (1985).

[13] R.W. Fredricks and L. Knopoff, The reflection of Rayleigh waves by a high impedance obstacle on a half space. Geophysics, 25(6), 1195-1202 (1960).

[14] I.A. Viktorov, Rayleigh and Lamb Waves: Physical Theory and Applications, Plenum Press, New York (1967).